\documentclass[12pt]{iopart}
\usepackage{graphicx}

\newcommand{\avg}[1]{\langle #1 \rangle}

\begin{document}
\title[Asymptotic behaviour of lattice polygons with fixed area and varying perimeter]{Asymptotic behaviour of convex and column-convex lattice polygons with fixed area and varying perimeter}
\author{Mithun K Mitra$^1$, Gautam I Menon$^2$ and R Rajesh$^3$
}

\address{$^1$ Polymer Science and Engineering, 
              University of Massachusetts, Amherst 01003, USA}
\ead{mithun@polysci.umass.edu}
\address{$^2$The Institute of Mathematical Sciences,
              C.I.T. Campus, Taramani, Chennai 600113, India}
\ead{menon@imsc.res.in}
\address{$^3$The Institute of Mathematical Sciences,
              C.I.T. Campus, Taramani, Chennai 600113, India}
\ead{rrajesh@imsc.res.in}

\date{\today}

\begin{abstract}

We study the inflated phase of two dimensional lattice polygons, both convex
and column-convex, with fixed area $A$ and variable perimeter, when
a weight $\mu^t \exp[- Jb]$ is associated to a polygon with perimeter $t$ and 
$b$ bends. The mean perimeter is calculated  as a function of the 
fugacity $\mu$ and the bending rigidity $J$. 
In the limit $\mu \rightarrow 0$, the mean perimeter has the 
asymptotic behaviour $\avg{t}/4 \sqrt{A} \simeq 1 - K(J)/(\ln \mu)^2
+ \mathcal{O} (\mu/ \ln \mu) $.
The constant $K(J)$ is found to be the same for both types of polygons,
suggesting that self-avoiding polygons should also exhibit the same asymptotic 
behaviour. 
\end{abstract}

\vspace{2pc}

\maketitle

\section{\label{intro} Introduction}

The enumeration of lattice polygons weighted by area
and perimeter arises in many physical systems, including
vesicles \cite{leibler87,fisher91},
cell membranes \cite{satyanarayana04}, emulsions
\cite{faassen98}, polymers \cite{privman} and
percolation clusters \cite{rajesh05}. A central quantity of interest is the 
generating function
\begin{equation}
G(P,\mu,J)  = \sum_{A,t,b} C(A,t,b) e^{P A} \mu^t e^{-J b},
\end{equation}
where $C(A,t,b)$ is the number of self-avoiding polygons of area $A$,
perimeter $t$ and with $b$ bends. This is weighted by a pressure $P$
(conjugate to the area), a fugacity $\mu$ (conjugate to the perimeter) 
and a bending rigidity $J$ (conjugate to the number of bends).

Exact solutions exist for 
$G(P,\mu,0)$ when $C(A,t,b)$ is restricted to convex polygons 
\cite{lin91,bousquet92a,bousquet92b} or to column-convex polygons 
\cite{brak90a}. However, a general solution for self-avoiding polygons is 
unavailable and even the exact solutions for these restricted polygons
are complex enough that extracting
the asymptotics is non-trivial. The properties of self-avoiding polygons can be
studied by enumerating the number of configurations that correspond to a 
given area and perimeter. Exact enumeration results for self-avoiding polygons 
exist for values of the area $A$ up to $A=50$ and for all $t$ for these values 
of $A$ \cite{jensen03,jensensap}. The scaling function describing the scaling 
behaviour near the 
tricritical point $\lambda=0$ and $\mu=\kappa^{-1}$, where $\kappa$ is the 
growth constant for self-avoiding polygons, is also known exactly 
\cite{richard01,cardy01,richard02}.
A survey of different kinds
of lattice polygons and a review of related results
can be found in Refs.~\cite{bousquet96,rensburgbook}.

In an earlier paper \cite{mitra08}, we determined the
mean area of inflated convex and column-convex polygons of fixed
perimeter as a function of the pressure and bending rigidity. 
This case relates
to the problem of two-dimensional vesicles, or equivalently, pressurised
ring polymers \cite{leibler87,rudnick91,gaspari93,haleva06,mitra07} on a 
square lattice. We showed that in the limit of large pressure,
the expression for the average area was the same for convex and column-convex 
polygons. We also verified numerically that the same result held for 
the case of self-avoiding polygons.

In this paper, we consider the related problem of determining the mean
perimeter of a polygon of fixed area. 
We calculate the mean perimeter for convex 
[see Eq.~(\ref{convexasymp})] and column-convex 
[see Eqs.~(\ref{eq:ccp_fullt},\ref{columnconvexasymp})] polygons exactly. 
In the limit 
$\mu \rightarrow 0$, corresponding to inflated polygons, we show that for 
both convex and column-convex polygons, the perimeter is given by
\begin{equation}
\avg{t} = 4 \sqrt{A} \left[1 - 
\frac{1}{2 (\ln \mu)^2} \int_{1-\alpha}^1 dx \frac{\ln (1-x)}{x} \right] 
+ \mathcal{O}(\mu/\ln \mu) ,
\end{equation}
where $\alpha = e^{-2J}$. 
Since this result is the same for both convex and row-convex polygons, we
argue that this result should therefore also extend to the self-avoiding
case. 

In Sec.~\ref{sec2}, we outline the calculation scheme for determining the
mean perimeter. The results for convex polygons and column-convex polygons
are presented in Sec.~\ref{sec3a} and Sec.~\ref{sec3b} respectively.
In Sec.~\ref{sec4}, we generalise these results to the case of self avoiding
polygons and  compare the analytical results with results from exact
enumeration studies.

\section{\label{sec2} Outline of the calculation}

\begin {figure}
\includegraphics[width=0.48\columnwidth]{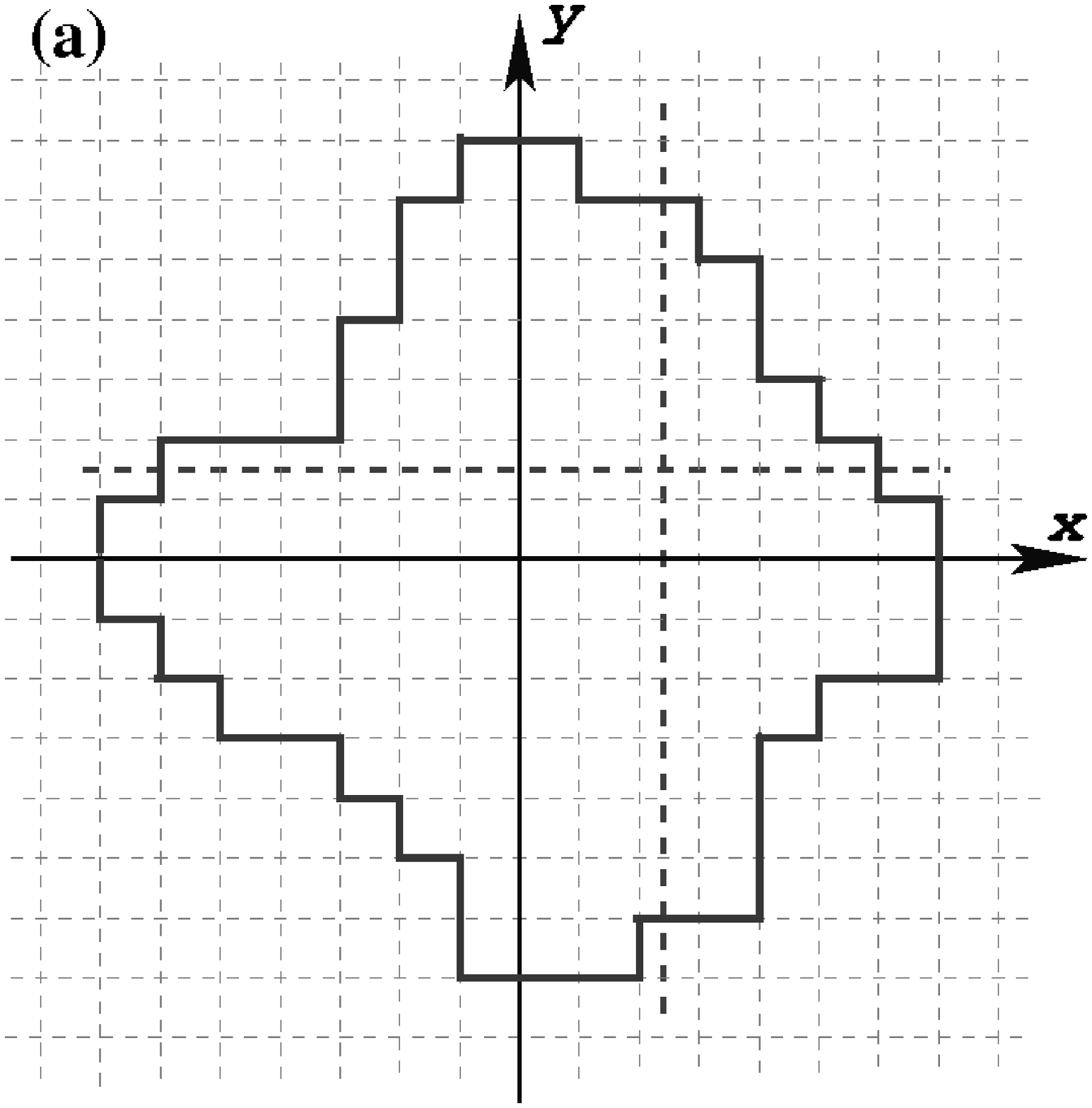}
\includegraphics[width=0.48\columnwidth]{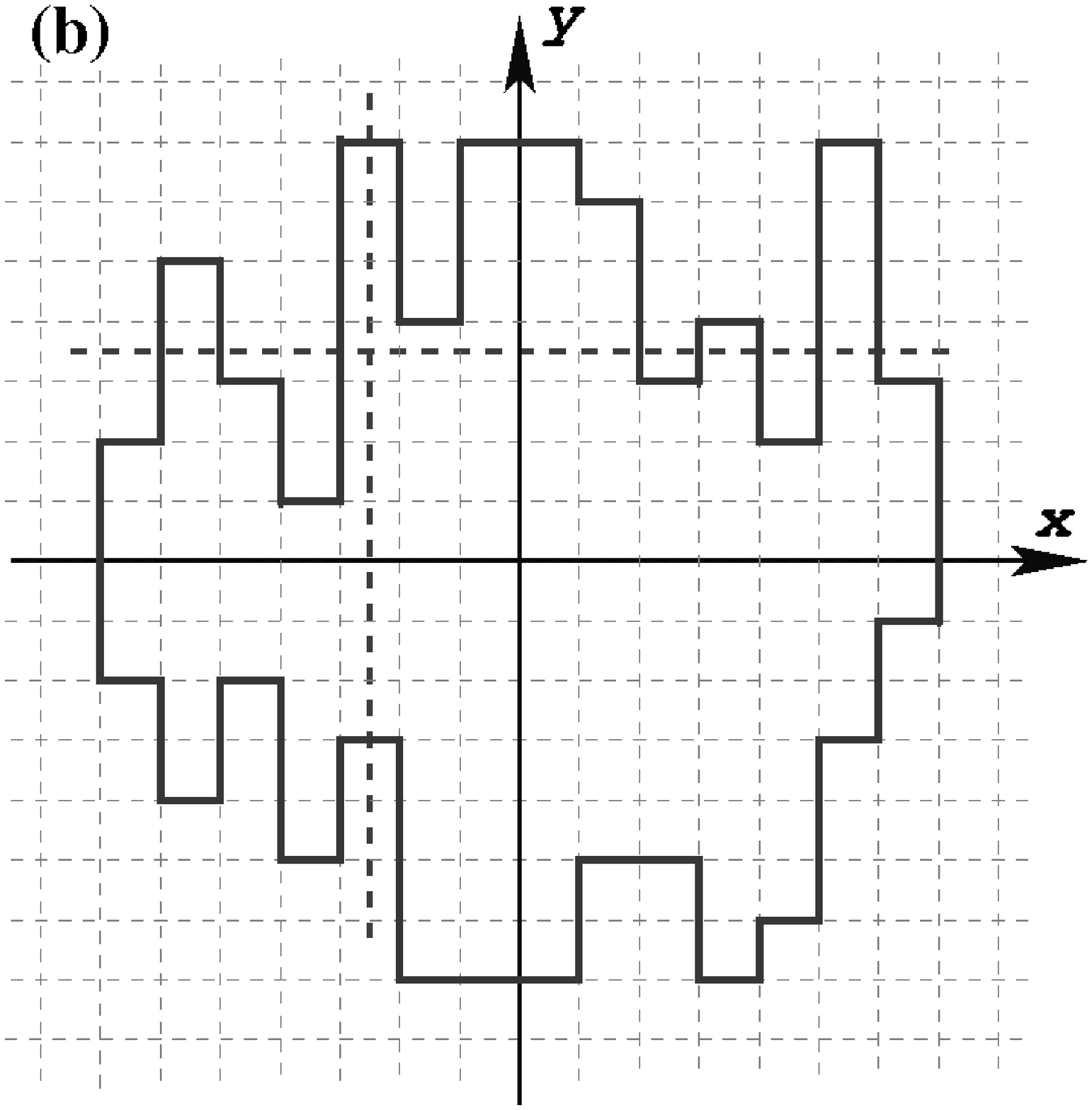}
\caption{\label{convexfig} 
(a) Schematic diagram of a convex polygon. 
Any vertical or horizontal line (thick dashed lines) 
intersects the convex polygon at either 
$0$ or $2$ points. (b) Schematic diagram of a column-convex polygon. Any
vertical line (thick dashed lines)
intersects the convex polygon at either
$0$ or $2$ points. }
\end {figure}

Convex polygons are those polygons that have exactly $0$ or $2$
intersections with any vertical or horizontal line drawn through the
midpoints of the edges of the lattice (Fig.~\ref{convexfig} (a)). 
Column-convex polygons 
are those polygons that have exactly $0$ or $2$ intersections with any 
vertical line drawn through the midpoints of the edges of the lattice. 
There is, however, no such restriction in the horizontal direction (Fig.~\ref{convexfig} (b)). Self avoiding polygons are polygons that have no
restrictions on overhangs.

\begin {figure}
\includegraphics[width=\columnwidth]{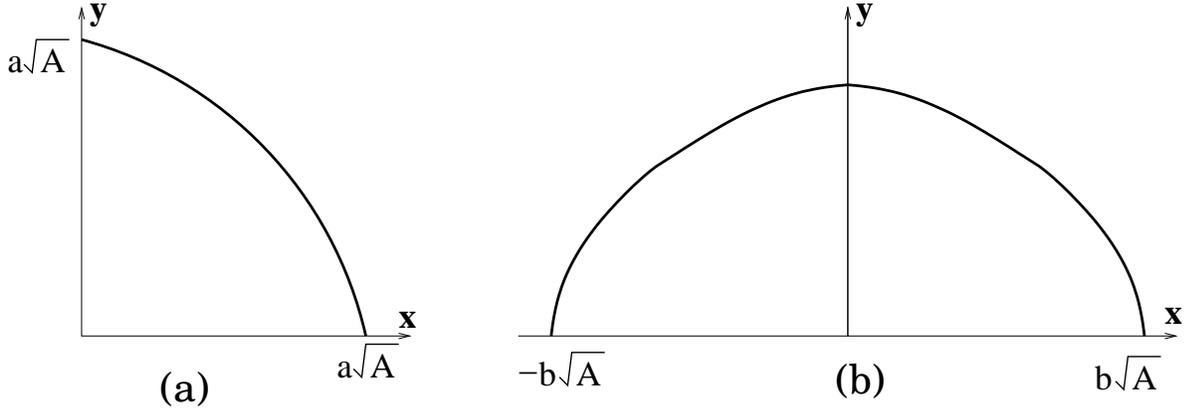}
\caption{\label{fig:shape} A schematic diagram of the equilibrium shape of
(a) convex polygon: the shapes in the other quadrants are obtained by
symmetry.
(b) column-convex polygons: the shape in the lower half plane is
obtained by reflection. }
\end {figure}

The shapes of convex and row-convex polygons are obtained by minimising
the free energy at fixed area, generalising the calculation presented in 
Ref.~\cite{mitra08}. The equilibrium shape of convex polygons is
invariant under rotations by $\pi/2$. Let the shape in the first
quadrant be represented by a curve $y_1(x)$ with endpoints at
$(a\sqrt{A},0)$ and $(0,a \sqrt{A})$, where $A$ is the area of the
polygon (see Fig.~\ref{fig:shape}(a)). In the case of column-convex polygons,
the equilibrium shape is invariant under reflection about the x-axis. 
Let $y_2(x)$ be the
shape of the column-convex polygon in the upper half plane with
endpoints at $(-b\sqrt{A},0)$ and $(b\sqrt{A},0)$ (see
Fig.~\ref{fig:shape}(b)). The free energy functionals for these shapes can
be written  as
\begin{eqnarray}
\mathcal{L}_1& =& \int_0^{a\sqrt{A}} \!\!\!\! dx \sigma_1(y_1')\sqrt{1+y_1'^2} - 
\frac{\lambda_1}{\sqrt{A}} \int_0^{a\sqrt{A}} \!\!\!\!  y_1 dx , 
		\label{lagconvex}
\end{eqnarray}
for convex polygons and
\begin{eqnarray}
\mathcal{L}_2& =& \int_{-b \sqrt{A}}^{b\sqrt{A}} \!\!\!\! 
dx \sigma_2(y_2') \sqrt{1+y_2'^2} - 
\frac{\lambda_2}{\sqrt{A}} \int_{-b\sqrt{A}}^{b\sqrt{A}} \!\!\!\!  y_2 dx ,
\label{lagcolumn}
\end{eqnarray}
for column-convex polygons.
The subscripts $1$ ($2$) denote convex (column-convex) polygons,
$\sigma $ is the free energy per unit length associated with a slope
$y'$ and $\lambda$'s are Lagrange multipliers. The Lagrange multipliers have
been scaled by $\sqrt{A}$ so that they become intensive quantities.

The angle dependent surface tension $\sigma$ was computed in \cite{mitra08}
using simple combinatorial arguments. For convex polygons $\sigma_1$ is given
by
\begin{equation}
\sigma_1(y') \sqrt{1+y'^2} = -f_1(\omega^*) - 2 \ln (\mu) ,
\label{convex_sigma}
\end{equation}
where
\begin{equation}
f_1(\omega) = y' \ln (\omega) + \ln [1-(1-\alpha)\omega] - \ln(1-\omega) ,
\end{equation}
and $\omega^*$ satisfies
\begin{equation}
\left. \frac{d f_1}{d \omega} \right|_{\omega^*} = 0 ,
\end{equation}
and $\alpha= e^{-2 J}$.

For column-convex polygons, the surface tension $\sigma_2$ is of the
form \cite{mitra08}
\begin{equation}
\sigma_2(y')  \sqrt{1+y'^2} = i B_0 y' -\ln \mu - \ln f_2(\mu,\alpha,B_0) ,
\label{column_sigma}
\end{equation}
where
\begin{equation}
f_2(\mu,\alpha,B) = \frac{1+(1-2\alpha)\mu^2 + \mu(\alpha-1)(e^{iB}+e^{-iB})}
{(1-\mu e^{iB})(1-\mu e^{-iB})} .
\end{equation}
and $B_0$ satisfies the equation,
\begin{equation}
\frac{d}{dB_0} \ln f_2(\mu,\alpha,B_0) = i y' .
\end{equation}
The equilibrium shape is obtained 
from Eqs.~(\ref{lagconvex}) and (\ref{lagcolumn})  by solving the
Euler Lagrange equation \cite{rottmannwortis84},
\begin{equation}
\frac{d}{d x} \frac{d}{d y'} \left[ \sigma_{1,2}(y') \sqrt{1+y'^2} \right] = 
-\frac{\lambda_{1,2}}{\sqrt{A}}.
\label{eq:euler}
\end{equation}

Using the above expressions for the surface tension $\sigma$, the
equilibrium shapes of convex and column-convex polygons were
computed exactly by solving the Euler Lagrange equations. 
We reproduce the final results, which will be the starting point of the
calculations presented in the next section. For convex polygons, the
equilibrium shape is given by
\begin{equation}
y_1 = \frac{\sqrt{A}}{\lambda_1} \ln \left[ \frac{1-c_1 e^{\lambda_1 x/\sqrt{A}}}{ 
\left[1-c_1 (1-\alpha)e^{\lambda_1 x/\sqrt{A}} \right]  c_1}  \right],
\label{eq:convexshape}
\end{equation}
with the constant $c_1 $ being given by,
\begin{equation}
c_1  = \frac{(1+e^{\lambda_1 a}) - \sqrt{(1+e^{\lambda_1 a})^2 - 4
e^{\lambda_1 a} (1-\alpha)}}
{2 (1-\alpha) e^{\lambda_1 a}} .
\label{eq:cc1}
\end{equation}

The equilibrium shape for column convex polygons is given by
\cite{mitra08}
\begin{eqnarray}
\label{columnconvexshapeeq}
\lefteqn{y_2(x) = -\frac{\sqrt{A}}{\lambda_2} \times  }     \\ \nonumber
&& \!\!  \ln \left[ \frac{(1-\mu e^{\lambda_2 x/\sqrt{A}}) 
(1- \mu e^{-\lambda_2 x/\sqrt{A}}) e^{-c_2}}
{1+(1-2\alpha)\mu^2 +\mu (\alpha-1)(e^{\lambda_2 x/\sqrt{A}}+
e^{-\lambda_2 x/\sqrt{A}})}\right] , 
\end{eqnarray}
where the constant $c_2$ is given by
\begin{equation}
c_2 = \ln \frac{(1-\mu g_2) (1- \mu g_2^{-1})}
{1+(1-2\alpha)\mu^2 +\mu (\alpha-1)(g_2+g_2^{-1})} ,
\label{ccconstant}
\end{equation}
where, $g_2 = \exp (\lambda_2 b)$. The constants $a$ and $b$ (see
Fig.~\ref{fig:shape})  are
chosen so that the free energy is minimised.

\section{\label{sec3} Results}

Starting from the equilibrium shapes [Eqs. (\ref{eq:convexshape}) and 
(\ref{columnconvexshapeeq})] described in Sec.~\ref{sec2}, we
now calculate the mean perimeter of convex and column-convex polygons
of fixed area $A$. 

\subsection{\label{sec3a} Convex polygons}

The Lagrange multiplier $\lambda_1$ is determined by the constraint
that the area under the curve is $A/4$: 
\begin{equation}
\int_0^{a\sqrt{A}} y_1 dx = \frac{A}{4} .
\end{equation}
Substituting for the form of the equilibrium curve as given by Eq.~(\ref{eq:convexshape}), we obtain,
\begin{equation}
\frac{\lambda_1^2}{4} = \int_0^{\ln g_1 } dz \ln \left[ 
\frac{1-c_1 e^z}{c_1  [1-c_1 (1-\alpha)e^z]} \right] ,
\label{eq:pt}
\end{equation}
where $g_1  = \exp (\lambda_1 a)$. 

The free energy of the equilibrium shape is obtained by substituting
the expressions for the equilibrium curve Eq.~(\ref{eq:convexshape})
and the surface tension Eq.~(\ref{convex_sigma}), into 
Eq.~(\ref{lagconvex}). Simplifying, we obtain
\begin{equation}
\mathcal{L}_1 = \frac{\lambda_1 \sqrt{A}}{2} + 2 a \sqrt{A} \ln c_1  - 2 \ln \mu~ a \sqrt{A} .
\end{equation}
The parameter $a$ is chosen to be that value that minimises the above
expression for the free energy, i.e. $a$ satisfies
$\partial{\mathcal{L}_1}/\partial{a} = 0$. This gives,
\begin{equation}
\frac{1}{2} \frac{d\lambda_1}{da} + 2 \ln c_1  + 
\frac{2 a}{c_1 }\frac{dc_1 }{da} - 2 \ln \mu = 0.
\end{equation}
To calculate the first term, we differentiate Eq.~(\ref{eq:pt})
with respect to $a$ to obtain
\begin{eqnarray}
\frac{\lambda_1}{2} \frac{d \lambda_1}{d a} &=& \frac{1}{c_1 } 
\frac{d c_1 }{d a} 
\ln\!\! \left[ \frac{(1-c_1  g_1 ) (1-c_1)^{-1} g_1^{-1}}{[1-c_1 (1-\alpha)g_1 ]
[1-c_1 (1-\alpha)]} \right],
\label{eq:dpda}
\end{eqnarray}
where the constant $c_1 $ can be expressed in terms of $g_1 $ as
\begin{equation}
c_1  = \frac{1+g_1 -\sqrt{(1+g_1 )^2-4(1-\alpha)g_1 }}{2 (1-\alpha) g_1 } .
\label{eq:cofg}
\end{equation}

Substituting for $d\lambda_1/da$ from Eq.~(\ref{eq:dpda}) and using 
Eq.~(\ref{eq:cofg}) and simplifying, we obtain the solution
\begin{equation}
c_1  = \mu .
\end{equation}
The expression for $g_1 $ for this value of $c_1 $ is obtained by
replacing $c_1 $ with $\mu$ in Eq.~(\ref{eq:cofg}) :
\begin{equation}
g_1  = \frac{1-\mu}{\mu+\mu^2(\alpha-1)}.
\end{equation}
Knowing $g_1 $, the parameter $a$ can  easily be obtained as
\begin{equation}
a = \frac{1}{\lambda_1} \ln g_1  = \frac{1}{\lambda_1} \ln 
\big[ \frac{1}{\mu(1+\mu(\alpha-1))} \big] .
\end{equation}
The value of the mean perimeter is equal to 
\begin{equation}
\avg{t} = 8 a\sqrt{A},
\end{equation}
where the factor $8$ accounts for all the four quadrants.

When $\mu \rightarrow 0$, the shape reduces to  a square and the perimeter
equals $4 \sqrt{A}$. We would like to find the corrections for small
values of $\mu$. For small $\mu$,
\begin{equation}
\ln g_1  = -\ln \mu - \mu (\alpha -1) + O(\mu^2) .
\end{equation}
Expanding the expression Eq.~(\ref{eq:pt}) for $\lambda_1$ for
small $\mu$, we obtain
\begin{equation}
\frac{\lambda_1^2}{4} = (\ln \mu)^2 \big[ 1 + \frac{1}{(\ln \mu)^2} 
\int_{1-\alpha}^1 dx \frac{\ln (1-x)}{x} \big] + 
O\left( \frac{\mu}{\ln \mu}\right).
\end{equation}
The parameter $a$ is then given by,
\begin{eqnarray}
a &=& \frac{1}{\lambda_1} \ln g_1   \nonumber \\
&=& \frac{1}{2} - \frac{1}{4 (\ln \mu)^2} \int_{1-\alpha}^1 dx \frac{\ln (1-x)}{x} 
+ O\left( \frac{\mu}{\ln \mu}\right).
\end{eqnarray}
The total perimeter $t = 8a\sqrt{A}$ is then given by,
\begin{equation}
\frac{\avg{t}}{4 \sqrt{A}} = 1 - 
\frac{1}{2 (\ln \mu)^2} \int_{1-\alpha}^1 dx \frac{\ln (1-x)}{x} 
+ O\left( \frac{\mu}{\ln \mu}\right).
\label{convexasymp}
\end{equation}
The results for the mean perimeter for convex polygons are shown in
Fig.~\ref{perimeterplots}.

\subsection{\label{sec3b} Column-convex polygons}

In this section, we calculate the mean perimeter of column-convex 
polygons of area $A$ for arbitrary $\mu > 0$. We start with Eqs.
(\ref{lagcolumn}), (\ref{column_sigma}), 
(\ref{columnconvexshapeeq}) and (\ref{ccconstant}).

The Lagrange multiplier $\lambda_2$ is fixed by the constraint,
\begin{equation}
\int_{-b \sqrt{A}}^{b \sqrt{A}} y_2 dx = \frac{A}{2} .
\end{equation}
On simplifying, this gives,
\begin{eqnarray}
\label{eq:ccp_pt}
\lefteqn{\frac{\lambda_2^2}{4} =} \\ \nonumber
&& \int_0^{\ln g_2} dz \left[ \ln \frac{(1-\mu e^z)(1-\mu e^{-z})}
{1+(1-2\alpha)\mu^2 +\mu (\alpha-1)(e^z+e^{-z})} -c_2 \right] ,
\end{eqnarray}
where $\ln g_2 = \lambda_2 b$.

The total free energy of the curve with ends fixed as $(-b\sqrt{A},0)$ and $(b\sqrt{A},0)$
can be calculated by substituting the equation for the equilibrium curve 
[Eq.~(\ref{columnconvexshapeeq})] into the expression for the 
Lagrangian [Eq.~(\ref{lagcolumn})]and simplifying,  thus yielding
\begin{equation}
\mathcal{L}_2 = \sqrt{A} [ \lambda_2 + 2 b (c_2 - \ln \mu)].
\label{eq:ccp_lag}
\end{equation}

The parameter $b$ is fixed by the condition that the free energy
should be a minimum, i.e. $d \mathcal{L}_2/d b = 0$ :
\begin{equation}
\frac{d\lambda_2}{db}+2 (c_2-\ln \mu) + 2 b \frac{dc_2}{db} = 0.
\label{eq:dldb}
\end{equation}
The first term is calculated by differentiating Eq. (\ref{eq:ccp_lag})
with respect to b, 
\begin{equation}
\frac{1}{2} \frac{d \lambda_2}{d b} = -b \frac{dg}{db} \frac{dc_2}{dg} .
\label{eq:dpdbeta}
\end{equation}

Substituting Eq. (\ref{eq:dpdbeta}) into Eq. (\ref{eq:dldb}) 
and simplifying, we obtain
\begin{equation}
c_2 = \ln \mu .
\end{equation}
This immediately allows the solution of $g_2$ by substituting for 
$c_2$ in Eq.~\ref{ccconstant},
\begin{eqnarray}
\label{eq:ccp_g}
g_2 &=& \frac{1-\mu+\mu^2+\mu^3(2 \alpha -1)}{2 \mu [1+(\alpha-1)\mu]} \\ 
&+& \frac{\sqrt{(1-\mu^2)(1-2\mu+2(1-2\alpha)\mu^3-(1-2\alpha)^2 \mu^4)}}
{2 \mu [1+(\alpha-1)\mu]} . \nonumber
\end{eqnarray}

The average perimeter is related to the chemical potential $\mu$ by the
relation,
\begin{equation}
\avg{t} = - 2 \mu \frac{\partial{\mathcal{L}_2}}{\partial{\mu}} ,
\end{equation}
where the factor of two accounts for the equilibrium shape in the lower 
half plane also. Substituting for the free energy, this implies,
\begin{equation}
\frac{\avg{t}}{4 \sqrt{A}} = -\frac{\mu}{2} \frac{\partial{\lambda_2}}{\partial{\mu}} .
\end{equation}
Now, using Eq.~\ref{eq:ccp_pt} for $\lambda_2$ and simplifying, we obtain,
\begin{eqnarray}
\label{eq:ccp_fullt}
\frac{\avg{t}}{4 \sqrt{A}} &=& -\frac{1}{\lambda_2} \ln 
\left[ \frac{1-\mu g_2}{g_2(g_2-\mu)}\right] \\
&-& \left( \frac{1+\mu^2 (2 \alpha -1)}{\lambda_2 \sqrt{k_1^2 - k_2^2}} 
\times \right. \nonumber \\
&& \left. \ln \left[ 
\frac{(k_1+k_2)(g_2+1)+\sqrt{k_1^2-k_2^2}(g_2-1)}{(k_1+k_2)(g_2+1)-
\sqrt{k_1^2-k_2^2}(g_2-1)} \right] 
\right) ,
\nonumber
\end{eqnarray}
where $k_1$ and $k_2$ are given by,
\begin{eqnarray}
k_1 &=& 1+(1-2\alpha)\mu^2, \\
k_2 &=& 2 \mu (\alpha-1) .
\end{eqnarray}

We would now like to determine the small $\mu$ behaviour of the mean
perimeter. As in the case of  the convex polygon, Eq.~(\ref{eq:ccp_g}) reduces to
\begin{equation}
\ln g_2 = -\ln \mu -\alpha \mu + \mathcal{O}(\mu^2) .
\end{equation}
In this limit, we can expand the right hand side of Eq.~(\ref{eq:ccp_fullt}) as
\begin{equation}
\frac{t}{4 \sqrt{A}} = \frac{1}{\lambda_2} [-2 \ln \mu - \alpha \mu + \mathcal{O} (\mu^2 \ln \mu)] .
\label{eq:ccp_t}
\end{equation}

Finally, we require the expansion of $\lambda_2$ in terms of $\mu$ and 
$\ln \mu$.
The equation for the Lagrange multiplier $\lambda_2$, Eq.~\ref{eq:ccp_pt}, can
be expanded as,
\begin{eqnarray}
\frac{\lambda_2^2}{4} &=& \int_{\mu}^{1-\alpha \mu} dx \frac{\ln(1-x)}{x}
+ \int_{\mu^2}^{\mu} dx \frac{\ln(1-x)}{x} -\ln g_2 \ln \mu \nonumber \\
&-& \int_0^{\ln g_2} dz \ln \left[ 1 + (1-2\alpha) \mu^2 
+\mu(\alpha -1) (e^z + e^{-z}) \right] , \nonumber \\
&&
\end{eqnarray}
where we have used the identities,
\begin{eqnarray}
\int_0^{\ln g_2} dz \ln (1-\mu e^z) &=& 
\int_{\mu}^{1-\alpha \mu} dx \frac{\ln(1-x)}{x} , \\
\int_0^{\ln g_2} dz \ln (1-\mu e^{-z}) &=& 
\int_{\mu^2}^{\mu} dx \frac{\ln(1-x)}{x} .
\end{eqnarray}
On further simplification, this yields, in the limit $\mu \rightarrow 0$,
\begin{equation}
\lambda_2 = 2 \ln \mu \left[ 1 +  
\frac{\int_{1-\alpha}^1 dx \ln(1-x) /x}{2(\ln \mu)^2} \right] .
\end{equation}
Substituting for $\lambda_2$ in Eq.~\ref{eq:ccp_t}, we obtain the average perimeter 
for column-convex polygons as,
\begin{equation}
\frac{\avg{t}}{4 \sqrt{A}} \simeq 1 - \frac{1}{2 (\ln \mu)^2} 
\int_{1-\alpha}^1 dx \frac{\ln (1-x)}{x} + \mathcal{O} (\mu/ \ln \mu) .
\label{columnconvexasymp}
\end{equation}
The column-convex polygon results, both for the general case and in the small
$\mu$ approximation, are shown in Fig.~\ref{perimeterplots}.

\section{\label{sec4} Self Avoiding Polygons}

In Sec.~\ref{sec3}, we calculated, as a function of the fugacity $\mu$ and
bending rigidity $J$, the
mean perimeter of convex and column-convex polygons of fixed area $A$.
For small $\mu$, the expression for the perimeter takes the form
\begin{equation}
\frac{\avg{t}}{4 \sqrt{A}} \simeq 1 - \frac{1}{2 (\ln \mu)^2} 
\int_{1-\alpha}^1 dx \frac{\ln (1-x)}{x} + \mathcal{O} (\mu/ \ln \mu),
\label{periasymp}
\end{equation}
for both convex and column-convex polygons. This is similar to the case 
of polygons with fixed perimeter, where the asymptotic expressions for
area match up to the second term as well \cite{mitra08}. 
Introducing overhangs in one direction to
convert convex polygons to column-convex polygons does not affect the second
term in the expansion Eq.~(\ref{periasymp}). It is therefore plausible 
that introducing overhangs in the vertical direction also 
does not affect the above expression, and hence that
the mean perimeter of self-avoiding polygons should also be described by the 
expression given in Eq.~(\ref{periasymp}), for small $\mu$.
\begin {figure}
\includegraphics[width=\columnwidth]{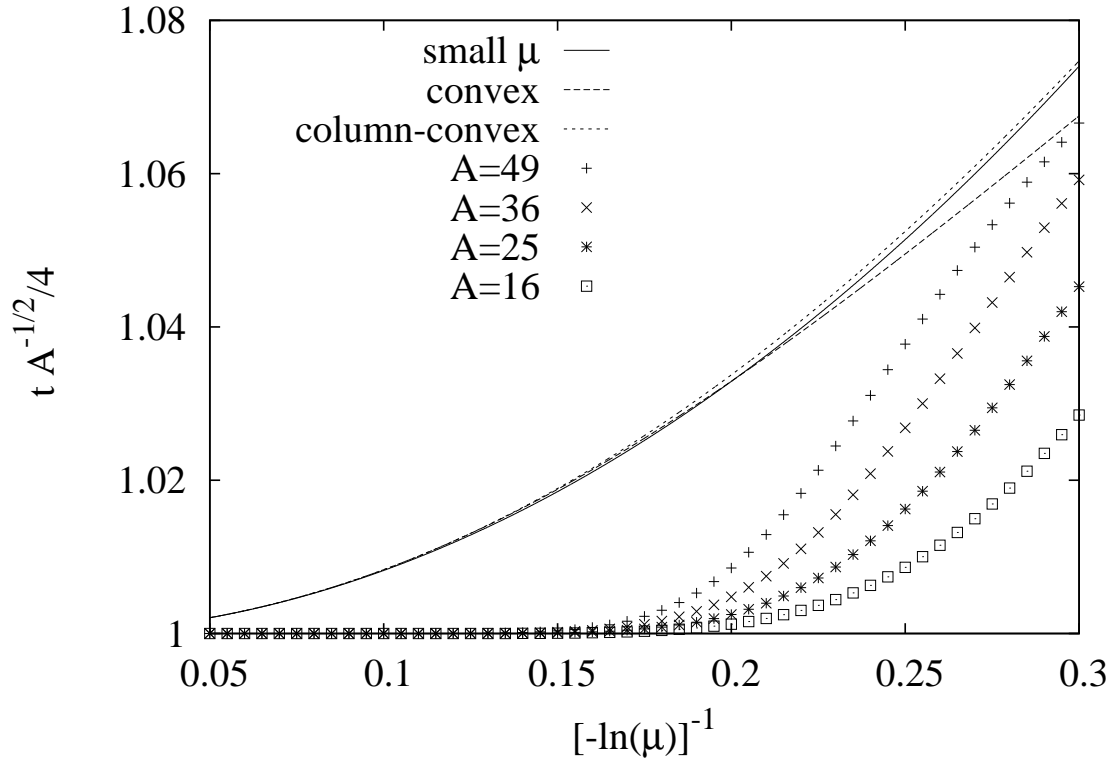}
\caption{\label{perimeterplots} 
The mean perimeter for convex and column-convex polygons are compared with
the small $\mu$ expansion result and data from exact enumeration of self
avoiding polygons of area $A$ up to $A=49$.
}
\end {figure}
 
In order to test this conjecture numerically, we used exact enumeration data
for polygons on the square lattice for the case $J=0$. For self-avoiding 
polygons, exact enumerations are available for areas up to $A=49$
\cite{jensensap}. The resulting plot for the average perimeter is shown in
Fig.~\ref{perimeterplots}. Unfortunately, with the data currently available
it is not 
possible to extrapolate to infinite $N$, preventing an unambiguous
test of this conjecture. Also, we know of  no simple Monte
Carlo algorithm that preserves area while varying perimeter, by which one
could access higher areas.

\section{\label{sec5} Summary and conclusions}

We now summarise the basic results of this paper. 
We  have  studied the variation of the perimeter, as a function of the chemical 
potential $\mu$ and bending rigidity $J$, for fixed area, 
for both convex and column-convex polygons. In each of these cases,  we 
have calculated the perimeter exactly. 
The asymptotic
behaviour in the limit $\mu \rightarrow 0$ 
coincides for both classes of polygons. We therefore
conjecture that overhangs are not important in the inflated regime, and hence
that self avoiding polygons should have the same asymptotic behaviour. 
It is important to obtain a numerical confirmation or rigorous proof of this
conjecture.

\end{document}